\documentclass[12pt,thmsb]{article}%
\usepackage{graphicx}
\usepackage{amsmath}
\usepackage{amsfonts}
\usepackage{amssymb}%
\begin{document}

\title{The recollapse problem of closed FRW models in higher-order gravity theories}
\author{John Miritzis\\Department of Marine Sciences, University of the Aegean\\Electronic address: imyr@aegean.gr}
\date{\today}
\maketitle

\begin{abstract}
We study the closed universe recollapse conjecture for positively curved FRW models
with a perfect fluid matter source and a scalar field which arises in the conformal
frame of the $R+\alpha R^{2}$ theory. By including ordinary matter, we extend the
analysis of a previous work. We analyze the structure of the resulted four-dimensional
dynamical system with the methods of the center manifold theory and the normal form
theory. It is shown that an initially expanding closed FRW universe, starting close to
the Minkowski spacetime, cannot avoid recollapse. We discuss the posibility that
potentials with a positive minimum may prevent the recollapse of closed universes.

\end{abstract}

\section{Introduction}

A closed Friedmann-Robertson-Walker (FRW) universe is often considered almost
synonymous to a recollapsing universe. This is mainly due to our experience
with the dust and radiation filled FRW models usually treated in textbooks.
That this picture is misleading follows clearly from an example found by
Barrow \emph{et al} \cite{bgt} according to which an expanding homogeneous and
isotropic model with spatial topology $S^{3}$ satisfying the weak, the strong,
the dominant energy conditions and the generic condition may not recollapse.
Thus the problem of recollapse of a closed universe to a second singularity is
delicate already in the FRW case.

The \emph{closed-universe recollapse conjecture} states roughly that a closed
universe cannot expand for ever, provided that the matter content satisfies
some energy condition and has non-negative pressures. The conjecture was found
true in certain spatially homogeneous cosmologies \cite{liwa}, in certain
spherically symmetric spacetimes \cite{burn} and in spacetimes admitting a
constant mean-curvature foliation that possesses a maximal hypersurface
\cite{rend}. In these investigations it has proved useful to demand that the
dominant energy condition and the positive pressure criterion hold (see also
Ref. \cite{hru} for a dynamical system approach).

In this paper we investigate the evolution of positively curved FRW models
with a scalar field having the potential which arises in the conformal frame
of the $R+\alpha R^{2}$ theory \cite{ba-co88,maed} and ordinary matter
described by a perfect fluid with energy density $\rho$ and pressure $p$. The
motivation for this choice was presented in \cite{miri1}. The purpose of the
present article is to generalize the results in \cite{miri1} by including
ordinary matter and to correct the mistake found therein.\footnote{In Ref.
\cite{miri1}, inequality (11) has the wrong direction (compare with
(\ref{ineq1}) in this paper). This mistake and a different rescalling (compare
with (\ref{resc1})) were the sources of the erroneous conclusion that an
initially expanding universe avoids recollapse. In fact, inequality (11) must
be reversed and as a consequence, the admissible trajectories of the system
(16) start below the line $H=\sqrt{2}r$ in FIG.1. This implies that an
initially expanded closed universe cannot avoid recollapse. Nevertheless, the
calculations and the mathematical analysis of the system (12) near the
equilibrium $\left(  0,0,0\right)  $ remain correct. Moreover, the above
mistake does not essentially affect the rest of the paper.}

The plan of the paper is as follows. In the next Section we write down the
field equations, as a constrained five-dimensional dynamical system. We use
the constraint equation to reduce the dimension of the system to four and
after a suitable change of variables the system becomes quadratic. In Section
III we analyze the structure of the equilibrium corresponding to the de Sitter
solution using the methods of the center manifold theory. Furthermore, we find
the so-called normal form of the dynamical system describing a large, slowly
expanding universe with low total energy density; we show that such a universe
cannot avoid recollapse. In the last Section, we consider potentials having a
strict positive minimum and argue that this class of potentials prevent a
closed universe from recollapse.

\section{Reduction to a 4-dimensional quadratic system}

In General Relativity the evolution of FRW models with a scalar field
(ordinary matter is described by a perfect fluid with energy density $\rho$
and pressure $p$) are governed by the Friedmann equation,
\begin{equation}
\left(  \frac{\dot{a}}{a}\right)  ^{2}+\frac{k}{a^{2}}=\frac{1}{3}\left(
\rho+\frac{1}{2}\dot{\phi}^{2}+V\left(  \phi\right)  \right)  , \label{fri1jm}%
\end{equation}
the Raychaudhuri equation,
\begin{equation}
\frac{\ddot{a}}{a}=-\frac{1}{6}\left(  \rho+3p+2\dot{\phi}^{2}-2V\right)  ,
\label{fri2jm}%
\end{equation}
the equation of motion of the scalar field,
\begin{equation}
\ddot{\phi}+3\frac{\dot{a}}{a}\dot{\phi}+V^{\prime}\left(  \phi\right)  =0,
\label{emsjm}%
\end{equation}
and the conservation equation,
\begin{equation}
\dot{\rho}+3\left(  \rho+p\right)  \frac{\dot{a}}{a}=0. \label{conssfjm}%
\end{equation}
We adopt the metric and curvature conventions of \cite{wael}. $a\left(
t\right)  $ is the scale factor, an overdot denotes differentiation with
respect to time $t,$ and units have been chosen so that $c=1=8\pi G.$ Here
$V\left(  \phi\right)  $ is the potential energy of the scalar field and
$V^{\prime}=dV/d\phi.$ We assume an equation of state of the form
$p=(\gamma-1)\rho,$ with $2/3<\gamma\leq2.$

In what follows we assume that the potential function of the scalar field is
\begin{equation}
V\left(  \phi\right)  =V_{\infty}\left(  1-e^{-\sqrt{2/3}\phi}\right)  ^{2}
\label{pote}%
\end{equation}
which arises in the conformal frame of the $R+\alpha R^{2}$ theory
\cite{maed}. \begin{figure}[h]
\begin{center}
\includegraphics[scale=0.7]{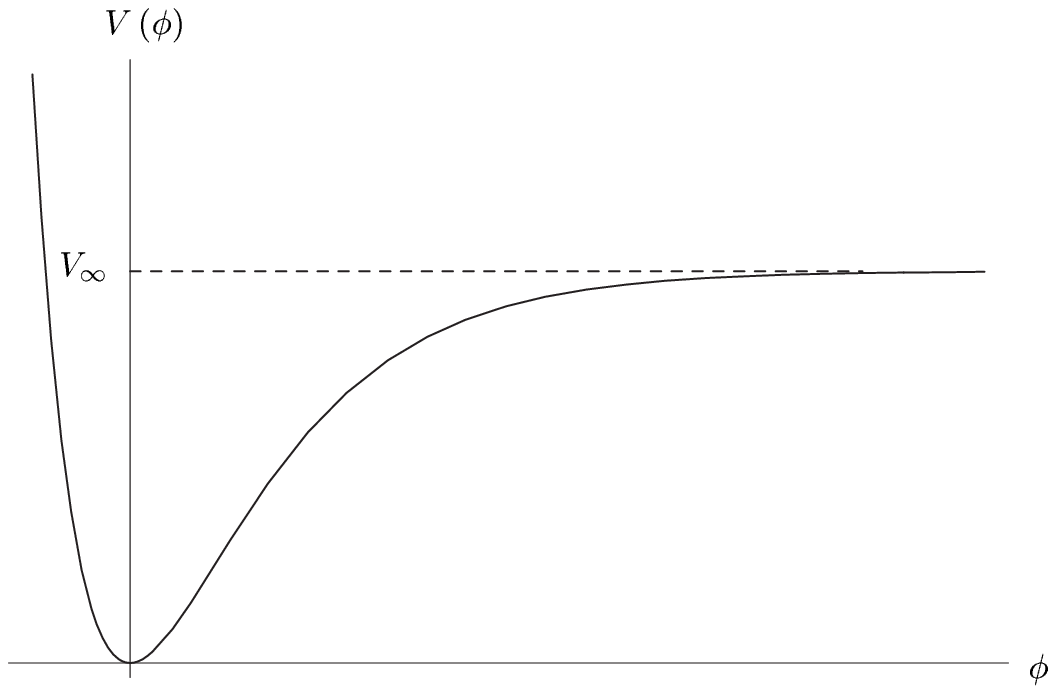}
\end{center}
\caption{The potential (\ref{pote})}%
\label{fig1}%
\end{figure}

The flat plateau of this potential is responsible for an early inflationary
period of the universe and, for homogeneous spacetimes provides a mechanism of
isotropization \cite{maed,comi}. From the field equations it is easy to see
that in an expanding universe the energy density of the scalar field, namely%
\[
E=\frac{1}{2}\dot{\phi}^{2}+V\left(  \phi\right)
\]
is a decreasing function of time. Since the energy density, $\rho,$ of
ordinary matter also decreases, it may happen that in a future time, $E$ be
comparable to $\rho.$ In particular, for closed, $k=1,$ models, once the scale
factor reaches its maximum value and recollapse commences i.e., $H<0,$ the
term $3H\dot{\phi}$ in (\ref{emsjm}) is no longer a damping factor, but acts
as a driving force which forces the field $\phi$ to oscillate with larger and
larger amplitude. If this be the case, the repulsive effect of the
cosmological term may drastically change the evolution of a classical FRW model.

Setting $\dot{\phi}=:y,\,\;\dot{a}/a=:H,$ we obtain from (\ref{fri1jm}) the constraint
equation
\begin{equation}
3H^{2}+3k/a^{2}=\rho+\frac{1}{2}y^{2}+V\left(  \phi\right)  , \label{constk}%
\end{equation}
which we use to eliminate $a$ from the evolution equations (\ref{fri2jm}%
)-(\ref{conssfjm}). As a consequence, the dimension of the dynamical system is
reduced to four and we obtain%
\begin{align}
\dot{\phi}  &  =y,\nonumber\\
\dot{y}  &  =-3Hy-V^{\prime}\left(  \phi\right)  ,\nonumber\\
\dot{\rho}  &  =-3\gamma\rho H,\nonumber\\
\dot{H}  &  =\frac{1}{3}V\left(  \phi\right)  -\frac{1}{3}y^{2}-\frac
{3\gamma-2}{6}\rho-H^{2}. \label{sys1}%
\end{align}
We remind the reader that the exponential potential which is popular in the
literature of scalar-field cosmologies has the nice property that $V^{\prime
}\propto V,$ which allows the introduction of normalized variables according
to the formalism of Wainwright \textit{et} \textit{al} \cite{wael}. For an
exponential potential the dimension of the dynamical system for a closed FRW
model reduces to three \cite{cogo}.

We simplify the system by rescaling the variables by the equations%
\begin{equation}
\phi\rightarrow\sqrt{3/2}\,\phi,\ \ y\rightarrow\sqrt{2V_{\infty}%
}\,\,y,\ \ \rho\rightarrow\frac{4V_{\infty}}{3}\rho,\ \ H\rightarrow
\sqrt{\frac{4V_{\infty}}{3}}\,H,\ \ t\rightarrow\sqrt{\frac{3}{4V_{\infty}}%
}\,t. \label{resc1}%
\end{equation}
Furthermore, in order to take account of the equilibrium point corresponding
to the point at \textquotedblleft infinity\textquotedblright\ and to remove
the transcendental functions, it is convenient to introduce the variable $u$
defined by
\begin{equation}
u:=e^{-\phi}, \label{u}%
\end{equation}
and system (\ref{sys1}) finally becomes
\begin{align}
\dot{u}  &  =-uy,\nonumber\\
\dot{y}  &  =-u+u^{2}-3Hy,\nonumber\\
\dot{\rho}  &  =-3\gamma\rho H,\nonumber\\
\dot{H}  &  =\frac{1}{4}\left(  1-u\right)  ^{2}-\frac{1}{2}y^{2}%
-\frac{3\gamma-2}{6}\rho-H^{2}. \label{sys3}%
\end{align}
Note that under the transformation (\ref{u}), the resulted four-dimensional
dynamical system (\ref{sys3}) is quadratic.

\textbf{Remark}: The system (\ref{sys3}) is not an arbitrary \textquotedblleft
free\textquotedblright\ four-dimensional system. In view of (\ref{constk}) the
initial conditions have to satisfy the condition $3H_{0}^{2}-\rho-\frac{1}%
{2}y_{0}^{2}-V\left(  \phi_{0}\right)  <0,$ or, in terms of the new
variables,
\begin{equation}
H_{0}^{2}-\frac{1}{3}\rho-\frac{1}{4}y_{0}^{2}-\frac{1}{4}\left(
1-u_{0}\right)  ^{2}<0. \label{ineq}%
\end{equation}
With a little manipulation of the equations (\ref{sys3}) it can be shown that,
once we start with initial conditions satisfying at time $t_{0}$ the
inequality (\ref{ineq}), the solutions of the system satisfy
\[
H\left(  t\right)  ^{2}-\frac{1}{3}\rho\left(  t\right)  -\frac{1}{4}y\left(
t\right)  ^{2}-\frac{1}{4}\left(  1-u\left(  t\right)  \right)  ^{2}<0
\]
for all $t>t_{0}.$ This is a general property of the Einstein equations,
namely that the subsequent evolution of the system is such that the solutions
respect the constraint. We conclude that the phase space of the system
(\ref{sys3}) is the set
\begin{equation}
\Sigma:=\left\{  \left(  u,y,\rho,H\right)  \in\mathbb{R}^{4}:H^{2}-\frac
{1}{3}\rho-\frac{1}{4}y^{2}-\frac{1}{4}\left(  1-u\right)  ^{2}<0\right\}  .
\label{ineq1}%
\end{equation}

\section{Stability analysis}

There are several equilibrium points of (\ref{sys3}). Some of them correspond
to static universes with a cosmological constant equal to $\sqrt{V_{\infty}}$.
In the study of the equilibrium points we note that $u=1$ corresponds to
$\phi=0,$ i.e., to the minimum of the potential and $u=0$ corresponds to
$\phi=\infty,$ i.e. to the flat plateau of the potential. In the following we
pay attention to the most interesting equilibrium solutions which are:

EQ1:\- $\left(  u=0,y=0,\rho=0,H=1/2\right)  .$ This corresponds to the de
Sitter universe with a cosmological constant equal to $\sqrt{V_{\infty}}$. We
analyze the flow of (\ref{sys3}) near EQ1 in the next subsection.

EQ2: $\left(  u=1,y=0,\rho=0,H=0\right)  .$ This corresponds to the limiting
state of an ever-expanding universe with $H\rightarrow0$ while the scalar
field approaches the minimum of the potential and the scale factor goes to
infinity. Equality in (\ref{ineq1}) which arises from the flat, $k=0,$ case
defines a set on the boundary of $\Sigma.$ We conclude that the point EQ2
which corresponds to the Minkowski solution, is located on this boundary. The
detailed structure of this equilibrium will be analyzed in subsection
\ref{normal}.

As we shall see, both equilibria are non-hyperbolic, i.e., some or all of the
eigenvalues of the Jacobian have zero real parts. That means that the
linearization theorem does not yield any information about the stability of
the equilibria and therefore, more powerful methods are needed. The study of
the qualitative behaviour of a dynamical system near a non-hyperbolic
equilibrium point is difficult even in two dimensions. There are two general
methods for simplifying a dynamical system having a non-hyperbolic
equilibrium. The first is the center manifold theory. According to the center
manifold theorem, the qualitative behaviour in a neighborhood of a
nonhyperbolic equilibrium point $\mathbf{q}$ is determined by its behaviour on
the center manifold near $\mathbf{q\,.}$ Since the dimension of the center
manifold is generally smaller than the dimension of the dynamical system, this
greatly simplifies the problem (cf. \cite{perko} and also \cite{rend1} for
cosmological applications). The second method is the normal form theory, which
consists in a nonlinear coordinate transformation that allows to simplify the
nonlinear part of the system (cf. \cite{perko} for a brief introduction). Both
methods are used in the next two subsections.

\subsection{Center manifold for the system at EQ1}

It is easy to see that at the equilibrium point $\mathbf{q}=$ $\left(
u=0,y=0,\rho=0,H=1/2\right)  ,$ the Jacobian matrix of (\ref{sys3}) has one
zero and three negative eigenvalues and, consequently the Hartman-Grobman
theorem does not apply. The center manifold theorem implies that there exists
a local 3-dimensional stable manifold through $\mathbf{q}$ (see for example
\cite{perko}). That means that all trajectories asymptotically approaching
$\mathbf{q}$ as $t\rightarrow\infty,$ lie on a 3-dimensional invariant
manifold. Since $\mathbf{q}$ is a non-hyperbolic fixed point, the topology of
the flow near $\mathbf{q}$ is non-trivial and is characterized by a
one-dimensional local center manifold containing $\mathbf{q}.$ We prove the
following result.\medskip

\textbf{Proposition:} \emph{The equilibrium point }$\mathbf{q}=\left(
0,0,0,1/2\right)  $\emph{ of (\ref{sys3}) is locally asymptotically
unstable.\medskip}

In order to determine the local center manifold of (\ref{sys3}) at
$\mathbf{q},$ we have to transform the system into a form suitable for the
application of the center manifold theorem. The procedure is fairly systematic
and will be accomplished in the following steps.

1. The Jacobian of (\ref{sys3}) at $\mathbf{q}=\left(  0,0,0,1/2\right)  $ has
eigenvalues $0,$ $-1,\ -3/2,$ and $-3\gamma/2$ with corresponding eigenvectors
$\left(  -2,4/3,0,1\right)  ^{^{T}},$ $\left(  0,0,0,1\right)  ^{^{T}},$
$\left(  0,1,0,0\right)  ^{^{T}}$ and $\left(  0,0,3,1\right)  ^{^{T}}.$ Let
$T$ be the matrix having as columns these eigenvectors. We shift the fixed
point to $\left(  0,0,0,0\right)  $ by setting $\widetilde{H}=H-1/2$ and write
(\ref{sys3}) in vector notation as
\begin{equation}
\mathbf{\dot{z}}=A\mathbf{z}+\mathbf{F}\left(  \mathbf{z}\right)  ,
\label{sys3a}%
\end{equation}
where $A$ is the linear part of the vector field and $\mathbf{F}\left(
\mathbf{0}\right)  =\mathbf{0}$.

2. Using the matrix $T$ which transforms the linear part of the vector field
into Jordan canonical form, we define new variables, $\left(  x,y_{1}%
,y_{2},y_{3}\right)  \equiv\mathbf{x}$, by the equations%
\begin{align*}
u  &  =-2x,\\
y  &  =\frac{4}{3}x+y_{2},\\
\rho &  =3y_{3},\\
\widetilde{H}  &  =x+y_{1}+y_{3},
\end{align*}
or in vector notation $\mathbf{z}=T\mathbf{x},$ so that (\ref{sys3a}) becomes
\[
\mathbf{\dot{x}}=T^{-1}AT\mathbf{x}+T^{-1}\mathbf{F}\left(  T\mathbf{x}%
\right)  .
\]
Denoting the canonical form of $A$ by $B$ we finally obtain the system
\begin{equation}
\mathbf{\dot{x}}=B\mathbf{x}+\mathbf{f}\left(  \mathbf{x}\right)  ,
\label{sys3b}%
\end{equation}
where $\mathbf{f}\left(  \mathbf{x}\right)  :=T^{-1}\mathbf{F}\left(
T\mathbf{x}\right)  .$ In components system (\ref{sys3b}) is%
\begin{align}
&  \left[
\begin{array}
[c]{c}%
\dot{x}\\
\dot{y}_{1}\\
\dot{y}_{2}\\
\dot{y}_{3}%
\end{array}
\right]  =\left[
\begin{array}
[c]{cccc}%
0 & 0 & 0 & 0\\
0 & -1 & 0 & 0\\
0 & 0 & -3/2 & 0\\
0 & 0 & 0 & -3\gamma/2
\end{array}
\right]  \left[
\begin{array}
[c]{c}%
x\\
y_{1}\\
y_{2}\\
y_{3}%
\end{array}
\right]  +\nonumber\\
&  \left[
\begin{array}
[c]{c}%
-\frac{4}{3}x^{2}-xy_{2}\\
\frac{4}{9}x^{2}-y_{1}^{2}-\frac{1}{2}y_{2}^{2}+\left(  3\gamma-1\right)
y_{3}^{2}-2xy_{1}-\frac{1}{3}xy_{2}+\left(  3\gamma-2\right)  xy_{3}+\left(
3\gamma-2\right)  y_{1}y_{3}\\
\frac{16}{9}x^{2}-\frac{3}{2}y_{2}^{2}-4xy_{1}-\frac{5}{3}xy_{2}%
-4xy_{3}-3y_{1}y_{2}-3y_{2}y_{3}\\
-3\gamma\left(  xy_{3}+y_{1}y_{3}+y_{3}^{2}\right)
\end{array}
\right]  . \label{sys3c}%
\end{align}

3. The system (\ref{sys3c}) is written in diagonal form%
\begin{align}
\dot{x}  &  =Cx+f\left(  x,\mathbf{y}\right) \nonumber\\
\mathbf{\dot{y}}  &  =P\mathbf{y}+\mathbf{g}\left(  x,\mathbf{y}\right)  ,
\label{sys3d}%
\end{align}
where $\left(  x,\mathbf{y}\right)  \in\mathbb{R}\times\mathbb{R}^{3},$ $C$ is
the zero $1\times1$ matrix, $P$ is a $3\times3$ matrix with negative
eigenvalues and $f,\mathbf{g}$ vanish at $\mathbf{0}$ and have vanishing
derivatives at $\mathbf{0.}$ The center manifold theorem asserts that there
exists a 1-dimensional invariant local center manifold $W^{c}\left(
\mathbf{0}\right)  $ of (\ref{sys3d}) tangent to the center subspace (the
$\mathbf{y}=\mathbf{0}$ space) at $\mathbf{0}.$ Moreover, $W^{c}\left(
\mathbf{0}\right)  $ can be represented as
\[
W^{c}\left(  \mathbf{0}\right)  =\left\{  \left(  x,\mathbf{y}\right)
\in\mathbb{R}\times\mathbb{R}^{3}:\mathbf{y}=\mathbf{h}\left(  x\right)
,\;\left\vert x\right\vert <\delta\right\}  ;\;\;\;\mathbf{h}\left(  0\right)
=\mathbf{0},\;D\mathbf{h}\left(  0\right)  =\mathbf{0},
\]
for $\delta$ sufficiently small (cf. \cite{perko}, p. 155). The restriction of
(\ref{sys3d}) to the center manifold is
\begin{equation}
\dot{x}=f\left(  x,\mathbf{h}\left(  x\right)  \right)  . \label{rest}%
\end{equation}
According to Theorem 3.2.2 in \cite{guho}, if the origin $x=0$ of (\ref{rest})
is stable (resp. unstable) then the origin of (\ref{sys3d}) is also stable
(resp. unstable). Therefore, we have to find the local center manifold, i.e.,
the problem reduces to the computation of $\mathbf{h}\left(  x\right)  .$

4. Substituting $\mathbf{y}=\mathbf{h}\left(  x\right)  $ in the second
component of (\ref{sys3d}) and using the chain rule, $\mathbf{\dot{y}%
}=D\mathbf{h}\left(  x\right)  \dot{x}$, one can show that the function
$\mathbf{h}\left(  x\right)  $ that defines the local center manifold
satisfies%
\begin{equation}
D\mathbf{h}\left(  x\right)  \left[  f\left(  x,\mathbf{h}\left(  x\right)
\right)  \right]  -P\mathbf{h}\left(  x\right)  -\mathbf{g}\left(
x,\mathbf{h}\left(  x\right)  \right)  =0. \label{h}%
\end{equation}
This condition allows for an approximation of $\mathbf{h}\left(  x\right)  $
by a Taylor series at $x=0.$ Since $\mathbf{h}\left(  0\right)  =\mathbf{0\ }%
$and $D\mathbf{h}\left(  0\right)  =\mathbf{0},$ it is obvious that
$\mathbf{h}\left(  x\right)  $ commences with quadratic terms. We substitute%
\[
\mathbf{h}\left(  x\right)  =:\left[
\begin{array}
[c]{c}%
h_{1}\left(  x\right) \\
h_{2}\left(  x\right) \\
h_{3}\left(  x\right)
\end{array}
\right]  =\left[
\begin{array}
[c]{c}%
a_{1}x^{2}+a_{2}x^{3}+O\left(  x^{4}\right) \\
b_{1}x^{2}+b_{2}x^{3}+O\left(  x^{4}\right) \\
c_{1}x^{2}+c_{2}x^{3}+O\left(  x^{4}\right)
\end{array}
\right]
\]
into (\ref{h}) and set the coefficients of like powers of $x$ equal to zero to
find the unknowns $a_{1},b_{1},c_{1},...$.

5. Since $y_{1}$ and $y_{3}$ are absent from the first of (\ref{sys3c}), we
give only the result for $h_{2}\left(  x\right)  .$ We find $b_{1}=32/27,$
$b_{2}=-32/81.$ Therefore, (\ref{rest}) yields%
\begin{equation}
\dot{x}=-\frac{4}{3}x^{2}-\frac{32}{27}x^{3}+\frac{32}{81}x^{4}+O\left(
x^{5}\right)  . \label{rest1}%
\end{equation}
It is obvious that the origin $x=0$ of (\ref{rest1}) is asymptotically
unstable (saddle point). The theorem mentioned after (\ref{rest}) implies that
the origin $\mathbf{x}=\mathbf{0}$ of the full four-dimensional system is
unstable. This completes the proof.

\subsection{Normal form of the system near EQ2\label{normal}}

Regarding the stability of this equilibrium, it is easy to see that the
eigenvalues of the Jacobian of (\ref{sys3}) are, $\pm i,0,0,$ i.e., it is
totally degenerate. Nevertheless, it is the most interesting case because in
other equilibria the scalar field reaches the flat plateau, which is
impossible if we restrict ourselves to initial values of $H$ smaller than
$\sqrt{V_{\infty}}$. We find the normal form of the system (\ref{sys3}) near
the equilibrium point $\left(  u=1,y=0,\rho=0,H=0\right)  .$ The idea of the
normal form theory is the following: Given a dynamical system with equilibrium
point at the origin, $\mathbf{\dot{x}}=A\mathbf{x}+\mathbf{f}\left(
\mathbf{x}\right)  ,$ where $A$ is the Jordan form of the linear part and
$\mathbf{f}\left(  \mathbf{0}\right)  =\mathbf{0}$, perform a non-linear
transformation $\mathbf{x}\rightarrow\mathbf{x}+\mathbf{h}\left(
\mathbf{x}\right)  ,$ where $\mathbf{h}\left(  \mathbf{x}\right)  =O\left(
\left\vert \mathbf{x}\right\vert ^{2}\right)  $ as $\left\vert \mathbf{x}%
\right\vert \rightarrow0,$ such that the system becomes \textquotedblleft as
simple as possible\textquotedblright.

To write the system in a form suitable for the application of the normal form
theory, we shift the fixed point to $\left(  0,0,0,0\right)  $ by setting
$x=u-1$ and the system becomes%
\begin{align}
\dot{x}  &  =-y-xy,\nonumber\\
\dot{y}  &  =x+x^{2}-3Hy,\nonumber\\
\dot{\rho}  &  =-3\gamma\rho H,\nonumber\\
\dot{H}  &  =\frac{1}{4}x^{2}-\frac{1}{2}y^{2}-\frac{3\gamma-2}{6}\rho-H^{2}
\label{sys4}%
\end{align}
Now we make the non-linear change of variables%
\begin{align*}
x  &  \rightarrow x-y^{2}-\frac{3\gamma-2}{16}\rho x+\frac{3}{4}Hy,\\
y  &  \rightarrow y+xy+\frac{3\gamma-2}{16}\rho y+\frac{3}{4}Hx,\\
\rho &  \rightarrow\rho,\\
H  &  \rightarrow H+\frac{3}{8}xy,
\end{align*}
and keeping only terms up to second order, we obtain the normal form of the
system, viz.,
\begin{align}
\dot{x}  &  =-y-\frac{3}{2}Hx,\nonumber\\
\dot{y}  &  =x-\frac{3}{2}Hy,\nonumber\\
\dot{\rho}  &  =-3\gamma\rho H,\nonumber\\
\dot{H}  &  =-\frac{3\gamma-2}{6}\rho-\frac{1}{8}\left(  x^{2}+y^{2}\right)
-H^{2}. \label{normal1}%
\end{align}
Note that the results are valid only near the origin.

Passing to cylindrical coordinates $\left(  x=r\cos\theta,y=r\sin\theta
,\rho=\rho,H=H\right)  ,$ we have
\begin{align}
\dot{r}  &  =-\frac{3}{2}rH,\;\nonumber\\
\;\dot{\theta}  &  =1,\nonumber\\
\dot{\rho}  &  =-3\gamma\rho H,\nonumber\\
\dot{H}  &  =-\frac{1}{8}r^{2}-H^{2}\;-\frac{3\gamma-2}{6}\rho.
\label{normal4d}%
\end{align}
We note that the $\theta$ dependence of the vector field has been eliminated,
so that we can study the system in the $(r,\rho,H)$ space. The equation
$\dot{\theta}=1$ means that the trajectory in the $x-y$ plane spirals with
angular velocity $1.$ The constraint (cf. (\ref{ineq1}))
\[
H^{2}<\frac{1}{3}\rho+\frac{1}{4}y^{2}+\frac{1}{4}x^{2}%
\]
becomes%
\begin{equation}
H^{2}<\frac{1}{4}r^{2}+\frac{1}{3}\rho. \label{phase}%
\end{equation}
We observe that the first and third of (\ref{normal4d}) can be written as a
differential equation
\[
\frac{d\rho}{dr}=2\gamma\frac{\rho}{r},
\]
which has the general solution
\begin{equation}
\rho=Cr^{2\gamma},\ \ \ \ C>0. \label{rho}%
\end{equation}
Therefore, for $\gamma=1,$ we obtain from (\ref{normal4d})%
\begin{align*}
\dot{r}  &  =-\frac{3}{2}rH,\\
\dot{H}  &  =-\frac{1}{8}r^{2}-\frac{1}{6}Cr^{2}-H^{2},
\end{align*}
It is convenient to rescale $r$ by%
\begin{equation}
r\rightarrow\sqrt{\frac{24}{4C+3}}\,r, \label{resc}%
\end{equation}
so that the projection of (\ref{normal4d}) on the $r-H$ plane is
\begin{align}
\dot{r}  &  =-\frac{3}{2}rH,\;\nonumber\\
\dot{H}  &  =-r^{2}-H^{2}\;. \label{2dim}%
\end{align}
This system belongs to a family of systems studied in 1974 by Takens
\cite{takens}. Note that the constraint (\ref{phase}) becomes in the new
variables
\[
H^{2}<2r^{2}%
\]
and we conclude that the phase space of (\ref{2dim}) is given by
\begin{equation}
-\sqrt{2}r\leq H\leq\sqrt{2}r. \label{phase1}%
\end{equation}

\begin{figure}[th]
\begin{center}
\includegraphics[scale=0.7]{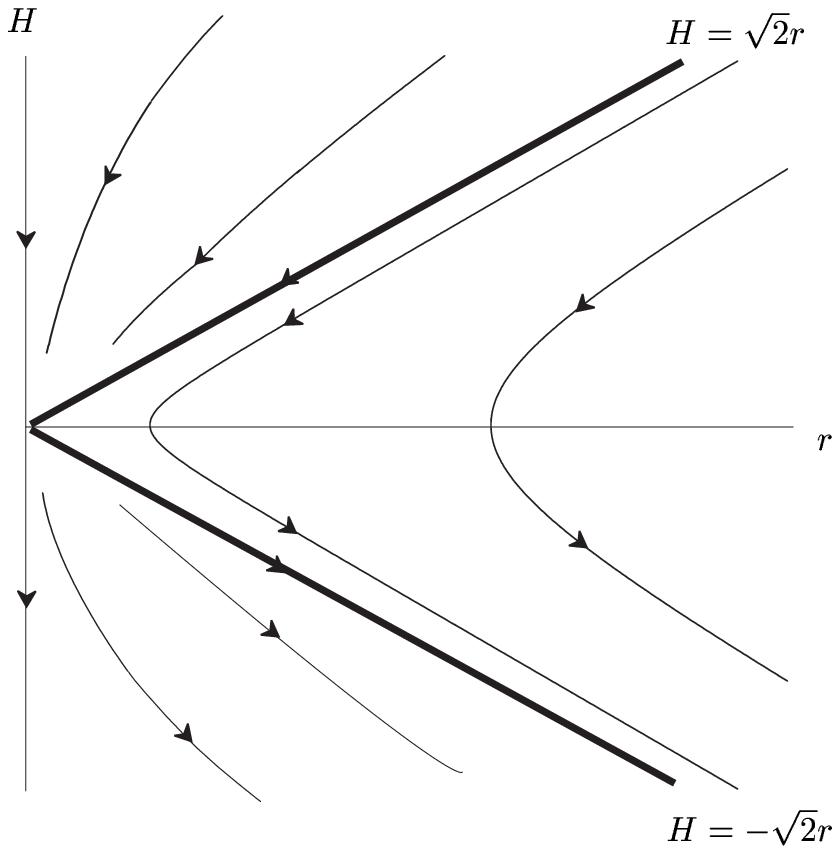}
\end{center}
\caption{The phase portrait of (\ref{2dim})}%
\label{fig2}%
\end{figure}

The phase portrait of (\ref{2dim}) is shown in FIG. 2 (see \cite{miri1} for a
detailed analysis). The system (\ref{2dim}) has invariant lines $H=cr$ with
$c=\pm\sqrt{2}.$ Since no trajectory can cross the line $H=cr,$ on any
trajectory starting in the first quadrant below the line $H=cr,$ $H$ becomes
zero at some time and the trajectory crosses vertically the $r-$axis. Once the
trajectory enters the second quadrant, $r$ increases and $H$ decreases. At
first sight, it seems probable that an initially expanding universe may avoid
recollapse; in fact all trajectories starting above the line $H=\sqrt{2}r$,
asymptotically approaches the origin and the corresponding universes would be
ever-expanding. But, (\ref{phase1}) implies that all trajectories with $H>0$
must start below the line $H=\sqrt{2}r.$ In conclusion, inequality
(\ref{phase1}) leaves no room for an ever-expanding closed universe, contrary
to what was claimed in \cite{miri1}.

We conclude that for an initially expanding universe $H$ continuously
decreases while $x$ and $y$ oscillate with decreasing amplitude. $H$ becomes
zero at some time and the scale factor reaches a maximum value. Subsequently
the universe begins to recollapse, i.e., $H$ continuously decreases below zero
while $x$ and $y$ oscillate with increasing amplitude. A typical trajectory of
(\ref{normal1}) is shown in FIG. 3, where the variable $\rho$ was suppressed.
One obtains qualitatively the same picture for all $\gamma\in\left[
2/3,2\right]  .$

\begin{figure}[h]
\begin{center}
\includegraphics[scale=0.8]{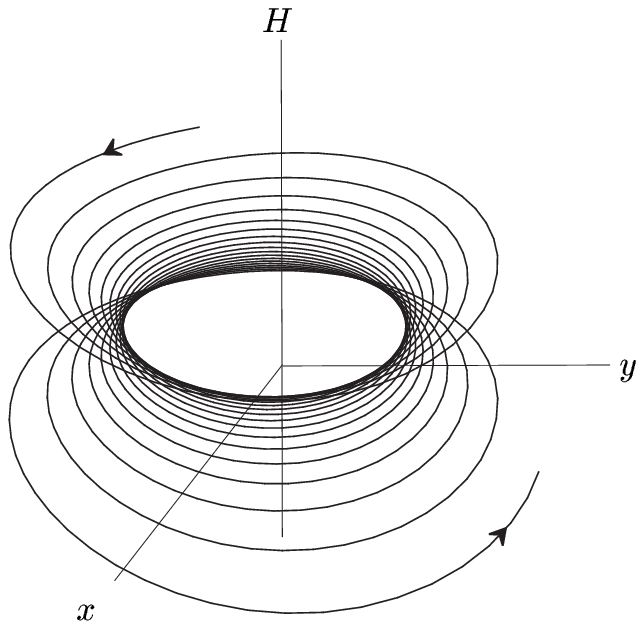}
\end{center}
\caption{A trajectory of (\ref{normal1})}%
\label{fig3}%
\end{figure}

\section{Further comments}

We have analyzed the qualitative behaviour of a positively curved FRW model
filled with ordinary matter and containing a scalar field with the potential
(\ref{pote}). This model is conformally equivalent to the positively curved
FRW spacetime in the simplest higher order gravity theory, namely the
$R+\alpha R^{2}$ theory. We have shown that even for large initial values of
$H$, near the flat plateau of the potential, the corresponding de Sitter
equilibrium is asymptotically unstable. Furthermore, an initially expanding
closed universe in the neighbourhood of EQ2 cannot avoid recollapse. For open
and flat models having potentials with a unique minimum, $V\left(  0\right)
=0,$ we have shown elsewhere \cite{miri}, that in an expanding universe, the
energy density $\rho$ of ordinary matter, the Hubble function $H$ and the
scalar field $\phi$ asymptotically approach zero. This theorem was proved
without referring to the precise form of the potential. Putting all these
results together, we may conjecture that potentials with a minimum equal to
zero, cannot provide a mechanism of late accelerating expansion of the
universe. On the other hand, in expanding universes with a potential having a
positive minimum, the scalar field rolls down to the minimum of the potential
and this residual cosmological term may explain the late accelerating
expansion of the universe \cite{rend2}.

We illustrate this idea by the example of the more general quadratic theory,
derived from the Lagrangian density $R+\alpha R^{2}-2\Lambda$. The
corresponding potential in the Einstein frame is
\begin{equation}
V_{\Lambda}\left(  \phi\right)  =V_{\infty}\left(  1-e^{-\sqrt{2/3}\phi
}\right)  ^{2}+\Lambda e^{-2\sqrt{2/3}\phi}. \label{pote1}%
\end{equation}
For every $\Lambda>0,$ the functions $V_{\Lambda}\left(  \phi\right)  $ have
the same qualitative behaviour as (\ref{pote}) but, have a positive minimum,
say $V_{\min}$ at some $\phi_{m}>0.$ Both $V_{\min}$ and $\phi_{m}$ increase
with increasing $\Lambda.$

We consider again expanding closed FRW models. It is easy to see that when
$V_{\Lambda}\left(  \phi\right)  =V_{\min},$ the system (\ref{sys1}) has an
equilibrium $\left(  \phi=\phi_{m},\ y=0,\ \rho=0,\ H=\sqrt{V_{\min}%
/3}\right)  $ representing the de Sitter solution. It can be shown simply by
the linearization theorem that this equilibrium is stable. To avoid
complicating expressions for the eigenvalues we proceed as in Section II and
obtain the following system (compare to (\ref{sys3}))%
\begin{align}
\dot{u}  &  =-uy,\nonumber\\
\dot{y}  &  =-u\left(  1-u\right)  +\lambda u^{2}-3Hy,\nonumber\\
\dot{\rho}  &  =-3\gamma\rho H,\nonumber\\
\dot{H}  &  =\frac{1}{4}\left(  1-u\right)  ^{2}+\frac{\lambda}{4}u^{2}%
-\frac{1}{2}y^{2}-\frac{3\gamma-2}{6}\rho-H^{2}, \label{sysla}%
\end{align}
with $\lambda=\Lambda/V_{\infty}.$ The equilibrium point
\[
\mathbf{p}=\left(  \phi=\frac{1}{1+\lambda},\ y=0,\ \rho=0,\ H=\frac{1}%
{2}\sqrt{\frac{\lambda}{1+\lambda}}\right)
\]
corresponds to the de Sitter solution with a cosmological term equal to
$V_{\min}.$ Linearization of (\ref{sysla}) near $\mathbf{p}$ is sufficient to
show that this point is a sink (all eigenvalues of the Jacobian matrix of
(\ref{sysla}) have negative real parts). Therefore $\mathbf{p}$ attracts all
nearby solutions and initially expanding closed universes enter a phase of
accelerating expansion. This attracting property of the de Sitter solution for
expanding models is well known from the cosmic no-hair conjecture and is not
restricted only to isotropic cosmology. We conclude that $\Lambda=0$ in
(\ref{pote1}) is a bifurcation value for closed models that recollapse or not.

However, de Sitter universe is not a global attractor for (\ref{sysla}).
Numerical experiments show that for highly curved models, or models filled
with an excess of ordinary matter, there are solutions of (\ref{sysla}) which
recollapse. Conditions to prevent the premature recollapse of closed models
were given in \cite{barr}.

Our results are based on the analysis of the behaviour of the dynamical system
(\ref{sys3}) near the equilibrium solutions. A rigorous proof of the closed
universe recollapse conjecture may come from the investigation of the global
structure of the solutions of (\ref{fri2jm})-(\ref{conssfjm}) with $k=+1$.{}
The study of the same question for Bianchi-IX models is an interesting
challenge for mathematical relativity.\textbf{\medskip}

\textbf{ACKNOWLEDGMENTS\medskip}

I thank N. Hadjisavvas and S. Cotsakis for fruitful discussions during the
preparation of this work.

\end{document}